\documentclass[journal,onecolumn]{IEEEtran}
\UseRawInputEncoding
\usepackage[ruled,linesnumbered]{algorithm2e}

\renewcommand{\IEEEQED}{\IEEEQEDclosed}
\usepackage{color}

\usepackage{subfigure}
\usepackage{graphicx,cite,epsfig,amssymb,amsmath,multirow,lettrine,flushend,extarrows}

\usepackage{threeparttable}

\begin{document}



\title{Deep Reinforcement Learning for Smart Grid Protection Against Coordinated Multistage Transmission Line Attacks}

\author{{Liang~Yu,~\IEEEmembership{Member,~IEEE}, Zhen Gao, Shuqi Qin, Meng Zhang, Chao Shen,~\IEEEmembership{Senior Member,~IEEE},\\ Xiaohong~Guan,~\IEEEmembership{Fellow,~IEEE}, Dong~Yue,~\IEEEmembership{Fellow,~IEEE}}
\thanks{\newline L. Yu is with Xi'an Jiaotong University, Xi'an 710049, China, and is also with Nanjing University of Posts and Telecommunications, Nanjing 210003, China. (email: liang.yu@njupt.edu.cn) \newline
Z. Gao, S. Qin and D. Yue are with Nanjing University of Posts and Telecommunications, Nanjing 210003, China. (email: zhen\_gao@126.com, shuqi\_qin@163.com, medongy@vip.163.com)\newline
M. Zhang, C. Shen, and X. Guan are with Systems Engineering Institute, Ministry of Education Key Lab for Intelligent Networks and Network Security, Xi'an Jiaotong University, Xi'an 710049, China. (email: mengzhang2009@xjtu.edu.cn, \{cshen,xhguan\}@sei.xjtu.edu.cn) \newline
}}


\maketitle

\begin{abstract}
With the increase of connectivity in power grid, a cascading failure may be triggered by the failure of a transmission line, which can lead to substantial economic losses and serious negative social impacts. Therefore, it is very important to identify the critical lines under various types of attacks that may initiate a cascading failure and deploy defense resources to protect them. Since coordinated multistage line attacks can lead to larger negative impacts compared with a single-stage attack or a multistage attack without coordination, this paper intends to identify the critical lines under coordinated multistage attacks that may initiate a cascading failure and deploy limited defense resources optimally. To this end, we first formulate a total generation loss maximization problem with the consideration of multiple attackers and multiple stages. Due to the large size of solution space, it is very challenging to solve the formulated problem. To overcome the challenge, we reformulate the problem as a Markov game and design its components, e.g., state, action, and reward. Next, we propose a scalable algorithm to solve the Markov game based on multi-agent deep reinforcement learning and prioritized experience replay, which can determine the optimal attacking line sequences. Then, we design a defense strategy to decide the optimal defense line set. Extensive simulation results show the effectiveness of the proposed algorithm and the designed defense strategy.
\end{abstract}


\section{Introduction}\label{s1}
With the increase of connectivity in power grid, fewer power outages are incurred since the high demand in a region can be supplied by the local and remote generators. However, such connectivity also brings threats to power grid\cite{Cheng2016}. To be specific, a large-scale power outage may be triggered by the failure of a critical component (e.g., a transmission line)\cite{Bialek2016}\cite{Jiang2019}. For example, over 80 percent of power in Pakistan has been lost due to the outage of a transmission line, which was caused by a physical sabotage\cite{Yan2015}. According to \cite{LiL2020}, a large-scale power outage can lead to substantial economic losses and serious negative social impacts. Therefore, it is of great importance to identify the critical transmission lines and deploy defense resources for their protection so that the negative impacts of power outages caused by intentional attacks or accidental damages could be reduced.

Many approaches have been proposed to identify the critical transmission lines in power grid under multiple outages, e.g., random chemistry search\cite{Eppstein2012}, graph theory\cite{Chu2017}, game theory\cite{Cheng2016}\cite{Farraj2016}\cite{Xiang2017}, bilevel programming\cite{Arroyo2010}, trilevel programming\cite{TianM2019}, stochastic programming\cite{Xiang2019}, and state failure network\cite{LiL2020}. However, the above efforts mainly focus on a single-stage attack (or one-shot atack), which means that attacking multiple elements at a time\cite{Ni2019}. Compared with a single-stage attack, a multistage or sequential attack (i.e., several attacks are launched in a time sequence) can lead to greater negative impacts for power grid\cite{Ni2019}. In \cite{Xiang2020}, a line defense method based on robust optimization and stochastic programming was proposed to minimize the load loss against multistage attacks under uncertainties (i.e., attacks have certain success probabilities). However, a possible cascading failure caused by multistage attacks was neglected. To identify the critical lines under multistage attacks that may initiate a cascading failure, Q-learning based methods were adopted in \cite{Ni2019}\cite{Yan2017}\cite{Wei2018}. When the number of transmission lines is large, storing value function using Q-table has a very high requirement on memory. To overcome this drawback, a particle swarm optimization (PSO) based heuristic approach was proposed in \cite{Jiang2019}. Although some advances have been made in the above efforts, they did not consider the problem of identifying the critical lines under coordinated multistage attacks (i.e., attacks are launched by multiple attackers coordinately and repeatedly until all attacking resources are used) that may initiate a cascading failure. Moreover, Q-learning and PSO based approaches have their respective limitations when the number of transmission lines is large. To be specific, Q-learning is known to be unstable or even to diverge when a nonlinear function approximator (e.g., a deep neural network) is used to
represent the value function\cite{Mnih2015} and PSO also has less stable performance\cite{Zhangzhidong2020}.

Based on the above observation, this paper intends to identify the critical lines under coordinated multistage attacks that may initiate a cascading failure and deploy limited defense resources optimally. To achieve the aim, we first formulate a total generation loss maximization problem with the consideration of multiple attackers and multiple stages. Due to the large size of solution space, it is very challenging to solve the formulated problem. To overcome the challenge, we reformulate the problem as a Markov game\cite{Littman1994}. Then, we propose an algorithm with low computational complexity to solve the Markov game based on multi-agent deep reinforcement learning (DRL) with attention mechanism\cite{LiangTSG2020} and prioritized experience replay\cite{Schaul2016}. Next, we design an optimal defense strategy based on the obtained optimal attacking line sequences and the number of defense resources (i.e., the number of lines can be protected\cite{Xiang2019}). Compared with existing works (e.g., Q-learning and PSO), the proposed DRL-based algorithm has a more stable performance\cite{Zhangzhidong2020}.

The contributions of this paper are summarized as follows.

\begin{itemize}
  \item We reformulate a total generation loss maximization problem in power grid under coordinated multistage line attacks based on the framework of Markov game and design its components, e.g., state, action, and reward.
  \item We propose a scalable algorithm to solve the Markov game based on multi-agent DRL with attention mechanism and prioritized experience replay, which can achieve higher generation loss by 21.46\%-85.98\% compared with existing schemes.
  \item We design an optimal defense strategy against coordinated multistage line attacks according to the feature of optimal attacking line sequences. When protecting 4.83\% of the total lines, the designed defense strategy can reduce generation loss by 6.62\%-17.79\% compared with other defense schemes.
\end{itemize}

The rest of this paper is organized as follows. In Section~\ref{s2}, we describe system model and formulate a total generation loss maximization problem as well as its variant. In Section~\ref{s3}, we propose an algorithm to solve the formulated problem. In Section~\ref{s4}, we design an optimal defense strategy against coordinated multistage attacks. In Section~\ref{s5}, performance evaluation is conducted. Finally, we draw a conclusion in Section~\ref{s6}.

\section{System Model and Problem Formulation}\label{s2}

\begin{figure}[!htb]
\centering
\includegraphics[scale=0.6]{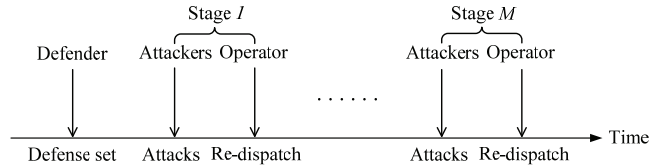}
\caption{The illustration of coordinated multistage attacks.}\label{fig_1}
\end{figure}

We consider a power system with $N$ transmission lines, which can be switched between in-service state and out-of-service state. When attacks are launched, transmission lines may become out of service\cite{Paul2017} and cascading failures in the power system may be triggered\cite{Eppstein2012}. In this paper, we mainly focus on coordinated multistage line switching attacks as shown in Fig.~\ref{fig_1}. To be specific, the pre-defined defense line set is first determined by the defender. Then, $K$ attacks are launched coordinately based on the observed line state information. Next, the power system operator will re-dispatch its components to reach a new steady state. Meanwhile, the quantity of generation loss is collected. In the second stage, $K$ attacks are launched again based on new transmission line state. The above process will repeat until the end of stage $M$. We assume that the objective of coordinated attackers is to maximize the total generation loss under the given defense line set and the number of attacking resources $MK$. Here, attacking resources are related to the number of attackers and the weapons/tools used when physical attacks are considered\cite{Xiang2019}. If cyber attacks are considered, attacking resources are associated with the capabilities and privileges of attackers\cite{Xiang2019}. In the following parts, we will first introduce the attack model and power system operator re-dispatching model. Then, we formulate a total generation loss maximization problem. Since it is difficult to solve the formulated problem, we reformulate the optimization problem as a Markov game.

\subsection{Attack Model}
Let $m_{l,t}$ be the state of transmission line $l$ at stage $t$. Then, we have
\begin{align} \label{f_1}
m_{l,t} = \left\{ \begin{array}{l}
1,\;\text{if~line $l$~is in-service at stage}~t,\\
0,\;\text{if~line $l$~is out-of-service at stage}~t.
\end{array} \right.
\end{align}

To describe the states of all transmission lines concisely, we define a line state vector as follows,
\begin{align} \label{f_2}
m_{t}=(m_{1,t},m_{2,t},\cdots,m_{N,t}).
\end{align}

Based on the current line state $m_t$, each attacker can choose attacking decision $b_{i,t}$ within a line set $\mathcal{B}_{i,t}$. If a unprotected line $l$ is being attacked by an attacker $i$ at slot $t$, its operational state will be changed. Thus, we have
\begin{align} \label{f_3}
m_{l,t+1} =0,\text{if}~m_{l,t}=1,~l=b_{i,t}~\text{and}~l \notin \mathcal{C},
\end{align}
where $\mathcal{C}$ denotes the pre-defined defense set.

\subsection{Power System Operator Re-dispatching Model}
The actions of attackers may trigger a power system cascading failure and the power system may be separated into several islands\cite{Ni2019}. To mitigate the cascading failure and achieve a new steady state, the power system operator can adjust branch flow by generator re-dispatching and load shedding with the consideration of ramping limits of generators\cite{Eppstein2012}\cite{Yang2019}. In this paper, the algorithms of mitigating cascading failures in \cite{Eppstein2012}\cite{Ni2019} are considered. Then, the dynamics of line states can be described by
\begin{align} \label{f_4}
m_{t + 1} = G_t(m_t,b_{1,t},b_{2,t},\cdots,b_{K,t},\mathcal{C}),
\end{align}
where $G_t(\cdot)$ denotes a line state transition function at slot $t$.

\subsection{Total Generation Loss Maximization Problem}
Based on the above models, a total generation loss maximization problem can be formulated as follows,
\begin{subequations}\label{f_5}
\begin{align}
(\textbf{P1})~&\max_{b_{i,t}\in \mathcal{B}_{i,t}}~\sum\limits_{t = 1}^M {{L_t}(s_t,b_{1,t},\cdots,b_{i,t},\cdots,b_{K,t},\mathcal{C})}  \\
s.t.&~\eqref{f_1}-\eqref{f_4},
\end{align}
\end{subequations}
where $L_t(\cdot)$ denotes the quantity of generation loss at slot $t$. Decision variables of \textbf{P1} are $b_{i,t}$ for all $i$ and $t$.

It is very challenging to solve the optimization problem \textbf{P1} due to the following reasons. Firstly, the size of search space is very large. When exhaustive search is adopted, the corresponding computational complexity is $O((\frac{N}{K})^{MK})$. If $K=3$, $N=186$, $M=3$, about $10^{16}$ simulations would be required to find the optimal attacking sequences, which is computational infeasible (e.g., about 317098 years are needed for completing all simulations if it takes one millisecond to run a simulation). Generally, random search methods (e.g., Monte Carlo simulation) can be adopted to avoid the search of the whole space. However, sufficient exploration of random search methods also results in high computational complexity due to the large size of search space. Secondly, the explicit expressions of $G_t$ and $L_t$ are unknown.

\subsection{Problem Reformulation}
Based on the above analysis, we intend to transform \textbf{P1} into another optimization problem. Since \textbf{P1} is a multi-player sequential decision-making problem under uncertainty, we reformulate \textbf{P1} as a Markov game\cite{Littman1994}, which is a multi-agent extension of Markov decision process. Since there are $K$ decision variables in \textbf{P1}, $K$ agents (i.e., $K$ attackers in this paper) are considered. According to \cite{LiangTSG2020}, a Markov game is defined by the following components, i.e., a set of environment (i.e., power system in this paper) states, $S$, a collection of action sets, $A_1$,~$\cdots$,~$A_{K}$, a state transition function, $F:~S\times A_1\times \ldots \times A_{K}\rightarrow \Pi(S)$, which defines the probability distribution over possible next states according to the given current state and actions for all agents, and a reward function for each agent $i$ ($1\leq i \leq K$), $R_i:~S\times A_1\times \ldots \times A_{K}\rightarrow \mathbb{R}$. At time slot $t$, agent $i$ takes action $a_{i,t}\in A_i$ based on its local observation $o_{i,t}\in \mathcal{O}_i$ (note that $o_{i,t}$ contains partial or whole information of the global state $s_t\in S$), with the aim of maximizing its expected return by learning a policy $\pi_i:~\mathcal{O}_i\rightarrow \Pi(A_i)$, which maps $o_{i,t}\in \mathcal{O}_i$ into a distribution over its set of actions. Here, the return is given by $\sum\nolimits_{j'=0}^{\infty}\gamma^{j'} r_{i,t+j'+1}(s_t,a_{1,t},\cdots,a_{K,t})$, where $r_{i,t}$ denotes the reward received by agent $i$ at slot $t$ and $\gamma \in [0,1]$ is used to determine that how much the policy favors immediate reward over long-term gain. Since the proposed algorithm in Section~\ref{s3} does not require a state transition function, we mainly focus on the design of state, action, and reward.

\subsubsection{State} In each time slot $t$, agent $i$ intends to attack a transmission line. Before launching an attack, it needs to know whether the transmission line is trapped or not. Therefore, the local observation of agent $i$ can be designed as follows, i.e., $o_{i,t}=s_t=m_t$. To describe local observations of all agents concisely, we define $o_t$ as follows, i.e., $o_t=(o_{1,t},o_{2,t},\cdots,o_{K,t})$.

\subsubsection{Action} According to the above description, the action of agent $i$ can be designed as follows, i.e., $a_{i,t}=b_{i,t}$. To describe actions of all agents concisely, we define $a_t$ as follows, i.e., $a_t=(a_{1,t},a_{2,t},\cdots,a_{K,t})$.

\subsubsection{Reward} To promote the coordinated attacks considered in \textbf{P1}, the same reward is assigned to all agents. Since the absolute value of generation loss is so large and not beneficial for the training of DRL agents, the percentage of generation loss $L_t^{'}=\frac{L_{t}}{L_{total}}$ at slot $t$ caused by coordinated attacks is adopted, where $L_{total}$ denotes the total generation output under normal operational state. Therefore, the reward of agent $i$ can be defined by $r_{1,t}=r_{2,t}=\dots=r_{N,t}=L_t^{'}$. To describe rewards of all agents concisely, we define $r_t$ as follows, i.e., $r_t=(r_{1,t},r_{2,t},\cdots,r_{K,t})$.

\section{The Proposed Algorithm}\label{s3}
In this section, we propose a solving algorithm for the Markov game problem based on multi-agent DRL with attention mechanism and prioritized experience replay\cite{Schaul2016}. To be specific, multi-agent-actor-critic (MAAC) framework in \cite{Iqbal2019} is used to learn policies of all agents efficiently, while prioritized experience replay can support high-efficient learning by replaying important transitions more frequently. In the following parts, we first introduce the basic principles of MAAC and prioritized experience replay. Then, the details of the proposed algorithm are provided. Finally, the computational complexity of the proposed algorithm is analyzed.

\subsection{MAAC Principle}

\begin{figure}[!htb]
\centering
\includegraphics[scale=0.6]{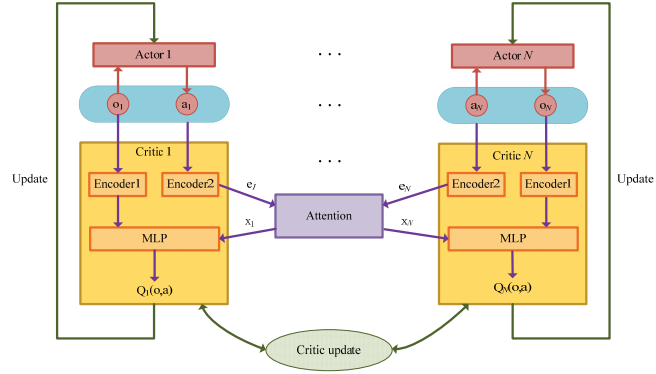}
\caption{The architecture of MAAC.}\label{fig_2}
\end{figure}

By adopting several techniques (e.g., soft actor-critic, attention mechanism, multi-task learning of critics, and multi-agent advantage function), MAAC has many advantages. For example, it is scalable to the number of agents. Moreover, it can train policies in environments with any reward setup and different action spaces for each agent. To illustrate the principle of MAAC more clearly, we provide the architecture of MAAC in Fig.~\ref{fig_2}, where actor module, critic module, encoder module, and attention module can be identified. Actor module takes action based on the local observation, while critic module evaluates the value of taken action at the given state. Encoder module can encode local observation (i.e., Encoder1) or the joint information of local observation and action (i.e., Encoder2). The outputs of Encoder2 modules in all critics will be taken as inputs of attention module, which outputs a value for the current agent and the value represents the contributions from other agents. Next, the value is used in computing the action-value function $Q_i^{\psi}(o,a)$ for agent $i$ (where $\psi$ is the weight parameter of critic network, $o=(o_1,o_2,\cdots,o_N)$, $a=(a_1,a_2,\cdots,a_N)$). To be specific, $Q_i^{\psi}(o,a)$ can be calculated by
\begin{equation}\label{f_6}
Q_i^{\psi}(o,a)=f_i(z_i(o_i),x_i),
\end{equation}
where $f_i$ is a two-layer multi-layer perceptron (MLP), $z_i$ is a one-layer MLP embedding function, and $x_i$ denotes the total contribution from other agents.

Let $q_i$ be a one-layer MLP embedding function and $e_i=q_i(o_i,a_i)$. Then, we have
\begin{equation}\label{f_7}
x_i=\sum\nolimits_{j\neq i}\kappa_j \Phi(W_ve_j),
\end{equation}
where $W_v$ is a shared matrix that transforms $e_j$ into a ``value", $\Phi$ is a non-linear activation function, $\kappa_j$ is the attention weight associated with agent $j$ and can be obtained as follows,
\begin{equation}\label{f_8}
\kappa_j=\text{exp}^{((W_ke_j)^TW_qe_i)}/\sum\nolimits_{\tau=1}^{N}\text{exp}^{((W_ke_{\tau})^TW_qe_i)},
\end{equation}
where $W_k$ and $W_q$ are shared matrixes that transform $e_j$ into a ``key" and transform $e_i$ into a ``query", respectively. Note that the above-mentioned ``value", ``key" and ``query" are similar to those in the key-value memory model\cite{Oh2016}.

Since some parameters (i.e., $W_k, W_q, W_v$) used in attention mechanism are shared by all agents, all critics are updated together by minimizing a joint regression loss function as follows,
\begin{equation}\label{f_9}
\mathcal{L}_Q(\psi)=\sum\nolimits_{i=1}^{K}\mathbb{E}_{(o,a,\tilde{o},r)\sim \mathcal{D}}[(Q_i^{\psi}(o,a)-y_i)^2],
\end{equation}
where $(o,a,\tilde{o},r)$ represents a tuple in replay buffer $\mathcal{D}$, $y_i=r_i(o,a)+\gamma \mathbb{E}_{\tilde{a}\in \pi_{\bar{\theta}}(\tilde{o})}[-\varphi\text{log}(\pi_{\bar{\theta_i}}(\tilde{a}_i|\tilde{o}_i))+Q_i^{\bar{\psi}}(\tilde{o},\tilde{a})]$, $\varphi$ is the temperature parameter in soft actor-critic and it determines the balance between maximizing entropy and maximizing reward, $\bar{\theta}$ is the weight parameter of target actor network.

Next, the weight parameter of actor network can be updated by policy gradient methods. To be specific, the gradient is given by
\begin{equation}\label{f_10}
\nabla_{\theta_i}J(\theta)=\mathbb{E}_{o\thicksim \mathcal{D}, a\thicksim \pi}[\nabla_{\theta_i}\text{log}(\pi_{\theta_i}(a_i|o_i))\rho_i(o_i,a_i)],
\end{equation}
where $\rho_i(o_i,a_i)=-\varphi\text{log}(\pi_{\theta_i}(a_i|o_i))+ Q_i^{\psi}(o,a)-d(o,a_{\backslash i})$, $\backslash i$ denotes the set of agents except $i$. Here, $Q_i^{\psi}(o,a)-d(o,a_{\backslash i})$ is called as the multi-agent advantage function, which can show that whether the current action will lead to an increase in expected return, where $d(o,a_{\backslash i})=\sum\nolimits_{\tilde{a}_i\in A_i}\pi_{\theta_i}(\tilde{a}_i|o_i)Q_i^{\psi}(o,(\tilde{a}_i,a_{\backslash i}))$.

\subsection{Prioritized Experience Replay}
Experience replay can help DRL agents to remember and reuse the past experiences, which improves the data efficiency and learning stability\cite{Mnih2015}. However, experience transitions in the traditional experience replay are uniformly sampled at random without considering their significance, resulting in low efficient learning. To improve this situation, the mechanism of prioritized experience replay\cite{Schaul2016} has been proposed to replay the important experience transitions more frequently and more effective learning can be achieved. To measure the significance of an experience, the magnitude of temporal-difference (TD) error is used. Since TD errors shrink slowly, initial high error transitions will be replayed more frequently. As a result, DRL agents will focus on a small subset of experiences and the approximation of value function will be over-fitted. To overcome the drawback, a stochastic sampling method is adopted\cite{Schaul2016}. Specifically, each transition is sampled according to the following probability, i.e.,
\begin{align}\label{f_11}
P_g=\frac{p_g^{\alpha}}{\sum\nolimits_{u}p_u^{\alpha}},
\end{align}
where $P_g$ is proportional to a transition's priority $p_g$ and parameter $\alpha$ represents how much prioritization is used. When $\alpha=0$, all experience transitions are sample uniformly.

To ensure that transitions with very small TD-errors still have chance of being sampled, $p_g$ can be configured as follows,
\begin{align}\label{f_12}
p_g=|\delta_g|+\epsilon,
\end{align}
where $\delta_g$ denotes TD-error related to transition $g$ and $\epsilon>0$.

Since prioritized experience replay may introduce estimation bias in approximating value function, importance-sampling weights are adopted to update the critic network as described in Algorithm~\ref{alg_1}.
\begin{align}\label{f_13}
\omega_g=(N_mP_g)^{-\beta},
\end{align}
where $N_m$ denotes the size of the replay buffer $\mathcal{D}$, parameter $0\leq\beta\leq 1$ is used to compensate for the non-uniform probability $P_g$. For stability reasons, all weights $\omega_g$ are scaled so that $\sum\nolimits_g\omega_g=1$. Typically, $\omega_g$ is computed by
\begin{align}\label{f_14}
\omega_g=(N_mP_g)^{-\beta}/\sum\nolimits_{g'}(N_mP_{g'})^{-\beta}.
\end{align}

\subsection{The Details of the Proposed Algorithm}
\begin{figure}[!htb]
\centering
\includegraphics[scale=0.48]{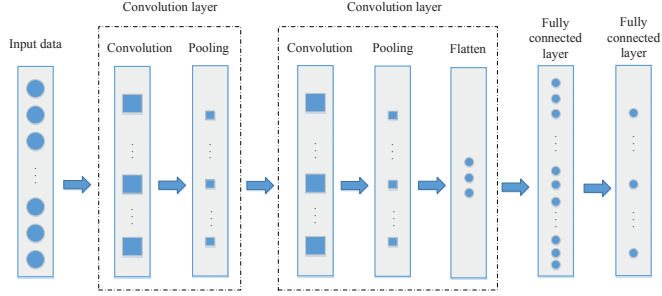}
\caption{The architecture of an actor network.}\label{fig_3}
\end{figure}

The proposed algorithm for the Markov game consists of two parts, i.e., training algorithm and execution algorithm. After the training process as shown in Algorithm~\ref{alg_1} is completed, the obtained policy represented by actor network can be used for execution. In execution algorithm, each attacker makes decision at time slot $t$ based on the observation state $o_{i,t}$ independently as shown in Algorithm~\ref{alg_2}. At the end of attacking stage, an optimal attacking sequence is obtained by each attacker. To reduce the state sensitivity of the learned policy, conventional neural networks are used in the representation of an actor network. To be specific, the architecture of actor network is shown in Fig.~\ref{fig_3}, where two convolutional layers and two fully connected layers can be identified. Each convolution layer consists of convolution operation and pooling operation. Note that pooling operation is adopted to reduce the dimensions of feature maps and increase the robustness of feature extraction. To support the connection between a convolution layer and a fully connected layer, an operation is adopted to flatten the multi-dimensional data into one-dimensional data at the end of the second convolution layer.

To better understand the training part of the proposed algorithm, we will explain it in more detail. In lines 1-3, memory replay buffer, weight parameters of all actor networks and critic networks, weight parameters of all target actor networks and target critic networks are initialized. In line 4, $Z$ episodes are considered and the length of each episode is $M$. In line 7, each agent observes environmental state $o_{i,t}$ and takes an action $a_{i,t}$ in parallel according to actor network $i$. In line 8, coordinated attacks are launched toward the smart grid and smart grid adjusts itself to achieve a new steady state. In line 9, each agent $i$ observes a new state $o_{i,t+1}$ and a reward $r_{i,t+1}$. In line 10, TD-error associated with each transition $(o_{i,t},a_{i,t},o_{i,t+1},r_{i,t})$ is calculated. Then, a total TD-error can be obtained for the joint experience transition $(o_t,a_t,o_{t+1},r_t)$. According to \eqref{f_12}, the priority $p_\ell$ for $\ell$th  joint experience transition can be obtained. Next, the tuple $(o_t,a_t,o_{t+1},r_t,p_\ell)$ is stored in the replay buffer. In line 12, when there are enough transitions for sampling, weight parameters of actor network and critic network will be updated every $T_{update}$ time slots. In line 13, $B_{size}$ transitions are sampled from the replay buffer according to a probability distribution in \eqref{f_11}. In lines 14-18, transition priority for experience transition $g$ is updated. In line 19, importance-sampling weight is calculated. In line 20, critic network is updated. Lines 21-23 are related to the update of actor networks. In line 24, target actor networks and target critic networks are updated.

\begin{algorithm}[h]
\caption{Training Algorithm}
\label{alg_1}
\setcounter{AlgoLine}{0}
\LinesNumbered
\KwIn{The numbers of attackers $K$ and stages $M$}
\KwOut{The weight parameter of actor network $\theta$}

Initialize experience replay buffer $\mathcal{D}$\;
Initialize weight parameters of actor network and critic network, i.e., $\theta$ and $\psi$\;
Initialize weight parameters of target actor network $\pi_i^{\bar{\theta}}$ and target critic network $Q_i^{\bar{\psi}}$ by copying: $\bar{\theta}\Leftarrow\theta$, $\bar{\psi}\Leftarrow\psi$\;

\For{$z$=0,~1,~$\cdots$,~$Z-1$}
{
   Initialize power system environment, and get initial observation state $o_{i,1}$ for each agent $i$\;

  \For{$t$=1,~2,~$\cdots$,~$M$}
  {
    Each agent $i$ selects action $a_{i,t}\sim \pi_i^{\theta}(\cdot|o_{i,t})$\;

    All agents launch attacks towards the power system according to the selected actions, and the power system adjusts itself to achieve a new steady state\;

    Each agent $i$ observes new state $o_{i,t+1}$ and receives reward $r_{i,t+1}$\;

    Calculate the priority $p_{\ell}$ for the $\ell\text{th}$ transition $(o_{t},a_{t},o_{t+1},r_{t+1})$ according to \eqref{f_12};

    Store transitions $(o_{t},a_{t},o_{t+1},r_{t+1},\bar{p}_{\ell})$ in the experience replay buffer\;

    \If{$\chi\geq KB_{\text{size}}$ and \text{mod}($zM+t$,$T_{\text{update}}$)=0}
      {
    Sample $B_{\text{size}}$ transitions $(o,a,\tilde{o},r$) from the experience replay buffer according to a probability distribution in \eqref{f_11};

    Calculate $Q_i^{\psi}(o_{i}^g,a_{i}^g)$ for all $i$ and $g$ ($1\leq g\leq B_{\text{size}}$)\;

    Calculate $\tilde{a}_{i}^g\sim \pi_i^{\bar{\theta}}(\tilde{o}_{i}^g)$ for all $i$ and $g$\;

    Calculate $Q_i^{\bar{\psi}}(\tilde{o}_{i}^g,\tilde{a}_{i}^g)$ for all $i$ and $g$\;

    Compute the total TD-error $\delta_g=\sum\nolimits_{i=1}^K (|y_i^g-Q_i^{\psi}(o_{i}^g,a_{i}^g)|)$;

    Update transition priority ${p_g}=\left| {{\delta_g}} \right|+\epsilon$;

    Compute importance-sampling weight $w_g$ according to \eqref{f_14};

    Update critic network by minimizing the weighted joint regressive loss function: $\mathcal{L}_Q(\psi)=\frac{\sum\nolimits_{i=1}^{K}\sum\nolimits_{g=1}^{B_{size}} \omega_g(Q_i^{\psi}(o_i^{g},a_i^{g})-y_i^{g})^2}{B_{size}}$\;

    Calculate $a_{i}^g \sim \pi_i^{\bar{\theta}}(o_{i}^g)$ for all $g$ and $i$\;

    Calculate $Q_i^{\psi}(o_{i}^g,~a_{i}^g)$ for all $g$ and $i$\;

    Update policies using \eqref{f_10}\;

    Update weight parameters of target actor network and target critic network:

    $\bar{\psi }\leftarrow \xi \psi+(1-\xi)\bar{\psi}$,~$\bar{\theta} \leftarrow \xi \theta+(1-\xi)\bar{\theta}$\;

  }

  }
}
\end{algorithm}

\begin{algorithm}[h]
\caption{Execution Algorithm}
\label{alg_2}
\setcounter{AlgoLine}{0}
\LinesNumbered
\KwIn{The weight parameter of actor network $\theta$, the numbers of attackers $K$ and stages $M$}
\KwOut{Coordinated attacking decisions $a_{t}$ ($1\leq t\leq M$)}

Each attacker $i$ ($1\leq i\leq K$) observes initial transmission line state vector $o_{i,1}$ in parallel\;

\For{$t$=1,~2,~$\cdots$,~$M$}
{

Each attacker $i$ selects its action $a_{i,t}$ in parallel according to the learned policy $\pi_{\theta}(\cdot|o_{i,t})$ at the beginning of slot $t$\;

Each attacker $i$ takes action $a_{i,t}$ in parallel, which affects the operation the power system\;

Each attacker $i$ receives new observation $o_{i,t+1}$ at the end of slot $t$\;

}
\end{algorithm}

\subsection{Computational Complexity of the Proposed Algorithm}
The computational complexity is a key performance metric for an algorithm. Firstly, we analyze the computational complexity of the execution algorithm. As shown in Fig.~\ref{fig_3}, two one-dimensional convolution layers and two fully connected layers are adopted in actor network. Since pooling operation intends to reduce the size of feature map of a convolution operation in the same convolution layer and a flattening operation has lower computational complexity than a pooling operation, the execution time of the proposed algorithm mainly depends on the computation complexities of two convolution operations and two fully connected operations. Since one-dimensional convolution is involved, the computational complexity of the $n$th convolution layer is $\mathcal{O}(X_{n} Y_{n} U_{n}^{in} U_{n}^{out})$, where $X_n$ is the kernel size, $Y_n$ is the size of feature map, $U_n^{in}$ and $U_n^{out}$ are the input channel number and the output channel number, respectively. Therefore, the computational complexity of all convolution layers is $\mathcal{O}(\sum_{n=1}^2 X_{n} {Y_n} U_{n}^{in} U_{n}^{out})$. Similarly, the computational complexity of all fully connected operations is $\mathcal{O}(\sum_{y=1}^2 V_{y}^{in} V_{y}^{out})$, where $V_{y}^{in}$ and $V_{y}^{out}$ represent the input and output size of $y$th fully connected layer, respectively. Since there are $M$ stages in the execution algorithm, the total computational complexity of the proposed algorithm at execution time is $\mathcal{O}_a=\mathcal{O}(M(\sum_{n=1}^2 {X_n} {Y_n} U_{n}^{in} U_{n}^{out} + \sum_{y=1}^2 V_{y}^{in} V_{y}^{out}))$.

Since three modules with one or two fully connected layers are adopted in a critic network, the computational complexity in the process of calculating value function can be represented by $\mathcal{O}_c=\mathcal{O}(E_{1,1}^{in} E_{1,1}^{out} + E_{2,1}^{in} E_{2,1}^{out} + E_{3,1}^{in} E_{3,1}^{out}+E_{3,2}^{in} E_{3,2}^{out})$, where $E_{\upsilon,w}^{in}$ and $E_{\upsilon,w}^{out}$ denote the input size and output size in $w$th fully connected layer of module $\upsilon$ ($\upsilon\in\{1,2,3\}$). In the training algorithm, its computational complexity mainly depends on lines 14-25. Since both actor network and critic network are involved in each learning with sampled $B_{size}$ transitions, the computational complexity of the training algorithm can be given by $\mathcal{O}(\frac{Z M B_{size}}{T_{update}} (\mathcal{O}_a + \mathcal{O}_c))$, where $\frac{ZM}{T_{update}}$ times of learning are conducted. It can be observed that the above computational complexity is independent of the number of agents $K$ since main steps in lines 14-25 can be conducted by each agent in parallel.

\section{The Designed Defense Strategy}\label{s4}
Based on the proposed algorithm, an optimal attacking line sequence related to each attacker can be found. Since we focus on the coordinated attacking effects (i.e., maximizing the total generation loss caused by all attackers in several stages), the optimal attacking line sequence related to each attacker is not necessarily unique. Therefore, multiple independent experiments should be conducted to identify more optimal attacking line sequences. Due to the large scale and high complexity of smart grid, it is impossible to protect all transmission lines, which requires enough financial and logistical support\cite{Wei2018}. Therefore, the lines with higher attacking frequencies should be protected with higher priorities under limited defense resources. The reason is that protecting such kinds of lines can make more optimal attacking sequences become non-optimal.

\begin{figure}[!htb]
\centering
\includegraphics[scale=0.55]{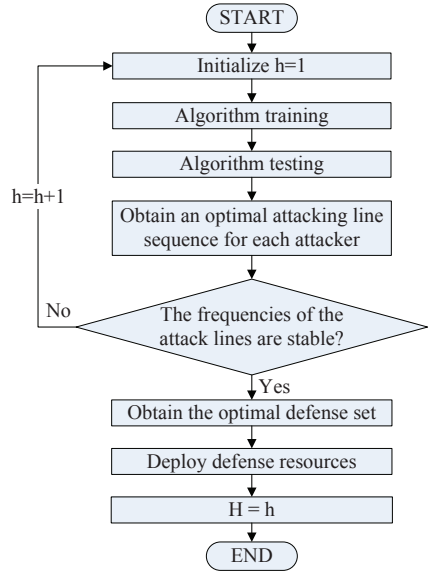}
\caption{The designed defense strategy.}\label{fig_4}
\end{figure}

To minimize the total generation loss caused by coordinated multistage attacks that may initiate a cascading failure, we design a defense strategy as shown in Fig.~\ref{fig_4}, where $h$ denotes the number of independent experiments. Let $\mathbb{A}_{i,h}$ be the optimal attacking line sequence of agent $i$ in experiment $h$. Next, for each element in a set $\mathbb{A}^{i}=\mathbb{A}_{i,1}\bigcup \mathbb{A}_{i,2}\bigcup \cdots \bigcup \mathbb{A}_{i,h}$, its occurrence frequency in $h$ experiments is collected. Then, the frequency of any line in the set $\mathbb{A}=\mathbb{A}^{1}\bigcup \mathbb{A}^{i} \bigcup \mathbb{A}^{K}$ can be obtained. When the frequencies of attack lines are stable (note that Euclidean distance between two frequency vectors can be used as the metric to describe the extent of stability), all lines are sorted in a descending order according to their frequencies. If the number of defense resources is $W$, $W$ lines with the highest frequencies would be selected in the optimal defense set. Finally, defense resources are deployed for line protection.

\section{Performance Evaluation}\label{s5}
In this section, we demonstrate the performances of the proposed algorithm and the designed defense strategy based on IEEE 118-bus system, which consists of 186 lines and 54 generators with total generation power 4242 MW. We first describe the simulation setup. Then, we introduce two benchmarks used for performance comparisons. Finally, we provide simulation results to illustrate the effectiveness of the proposed algorithm and the designed defense strategy.

\subsection{Simulation Setup}
To simulate the interaction process between multi-attacker and power system, a Python-based learning environment is created based on an open source code for cascading failure simulations in power systems\footnote{https://github.com/kevinzhou96/CascadingFailureSimulation}. To be specific, the open source code can simulate the propagation of branch failures in power systems under a variety of power redistribution rules. Note that all experiments are conducted based on a laptop with Intel CORE i7-9700 CPU and 16 GB RAM, and about 14.5 hours are needed to implement a complete experiment (including training and testing). In addition, main parameters used in the proposed algorithm the designed defense strategy can be found in Table~\ref{table_2}.


\begin{table}[!htb]
\renewcommand{\arraystretch}{1.3}
\caption{Main parameters} \label{table_2}\centering
\begin{tabular}{|c|c|}
\hline  Parameters                                                         &Value                 \\
\hline
\hline  Episode number                                                          &100000                \\
\hline       Memory size                                                        &15000                   \\
\hline       Batch size                                                         &128                   \\
\hline       Total number of runs                                               &10                   \\
\hline  Priority factor ($\alpha$)                                              &0.6                    \\
\hline  Constant in sampling probability ($\varepsilon$)                        &$1\times 10^{-6}$                    \\
\hline  Importance sampling weight correction coefficient ($\beta$)             &0.4                    \\
\hline  Importance sampling weight step-size                                    &0.001                    \\
\hline  The number of attacking resources                                       &9                      \\
\hline  The number of independent experiments                                   &15                     \\
\hline  The number of defense resources                                         &9                     \\
\hline
\end{tabular}
\end{table}

\subsection{Benchmarks}
For fair performance comparisons, two schemes that consider the same attacking resources as the proposed algorithm are adopted as follows,
\begin{itemize}
  \item \text{SAMS (Single-Agent-Multi-Stage)}: this scheme considers a single agent, which launches multistage attacks similar to \cite{Ni2019}. Different from \cite{Ni2019}, Deep Q-Networks (DQN) rather than Q-learning is adopted to train the agent since memory replay and target network are helpful to stabilize algorithmic performance.
  \item \text{MASS (Multiple-Agent-Single-Stage)}: this scheme considers coordinated single-stage attacking, i.e., coordinated attacks are launched by all attackers in one stage. Moreover, this scheme adopts the same algorithm as in this paper to train DRL agents. In other words, the proposed algorithm will be reduced to MASS if all attacking resources are used in the first stage.
\end{itemize}

\subsection{Simulation Results}
\subsubsection{Algorithmic Convergence Performance}
\begin{figure}[!htb]
\centering
\includegraphics[scale=0.45]{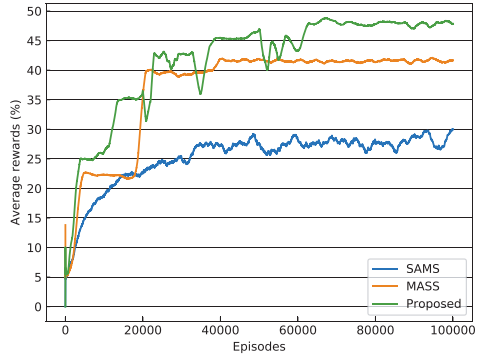}
\caption{Convergence performances of all schemes.}\label{fig_5}
\end{figure}

In Fig.~\ref{fig_5}, the average rewards over the past 200 episodes during the training process under different schemes are plotted. It can be seen that the proposed algorithm can achieve the best convergence performance. Moreover, MASS has better performance than SAMS. The reason is that MASS can support scalable learning with the help of attention mechanism and prioritized experience replay. In contrast, the DRL agent in SAMS has the largest action space but less efficient training method.

\subsubsection{Algorithmic Effectiveness}
\begin{figure}[!htb]
\centering
\includegraphics[scale=0.45]{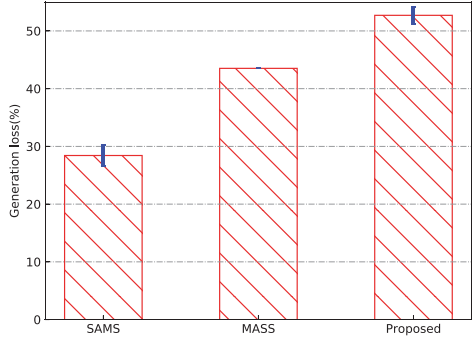}
\caption{Attacking performances under all schemes.}\label{fig_6}
\end{figure}

The average attacking performances under all schemes during the testing process are shown in Fig.~\ref{fig_6}, where 15 experiments and 95\% confidence interval are considered. It can be observed that the proposed algorithm achieves the best performance among all schemes. To be specific, the proposed algorithm can achieve higher generation loss by 85.98\% and 21.46\% compared with SAMS and MASS, respectively. The reason can be explained as follows. Compared with SAMS, the agent in the proposed algorithm has smaller action space and adopts prioritized experience replay for efficient exploration, resulting in better performance. Compared with MASS, the proposed algorithm can implement multistage attacking, which is helpful to learn better attacking policies if two or more optimal attacking lines are contained in the action space of an agent.

\subsubsection{The Performance of Defense Strategy}

\begin{figure*}
\centering
\subfigure[Line frequency stability]{
\begin{minipage}[b]{0.36\textwidth}
\includegraphics[width=1\textwidth]{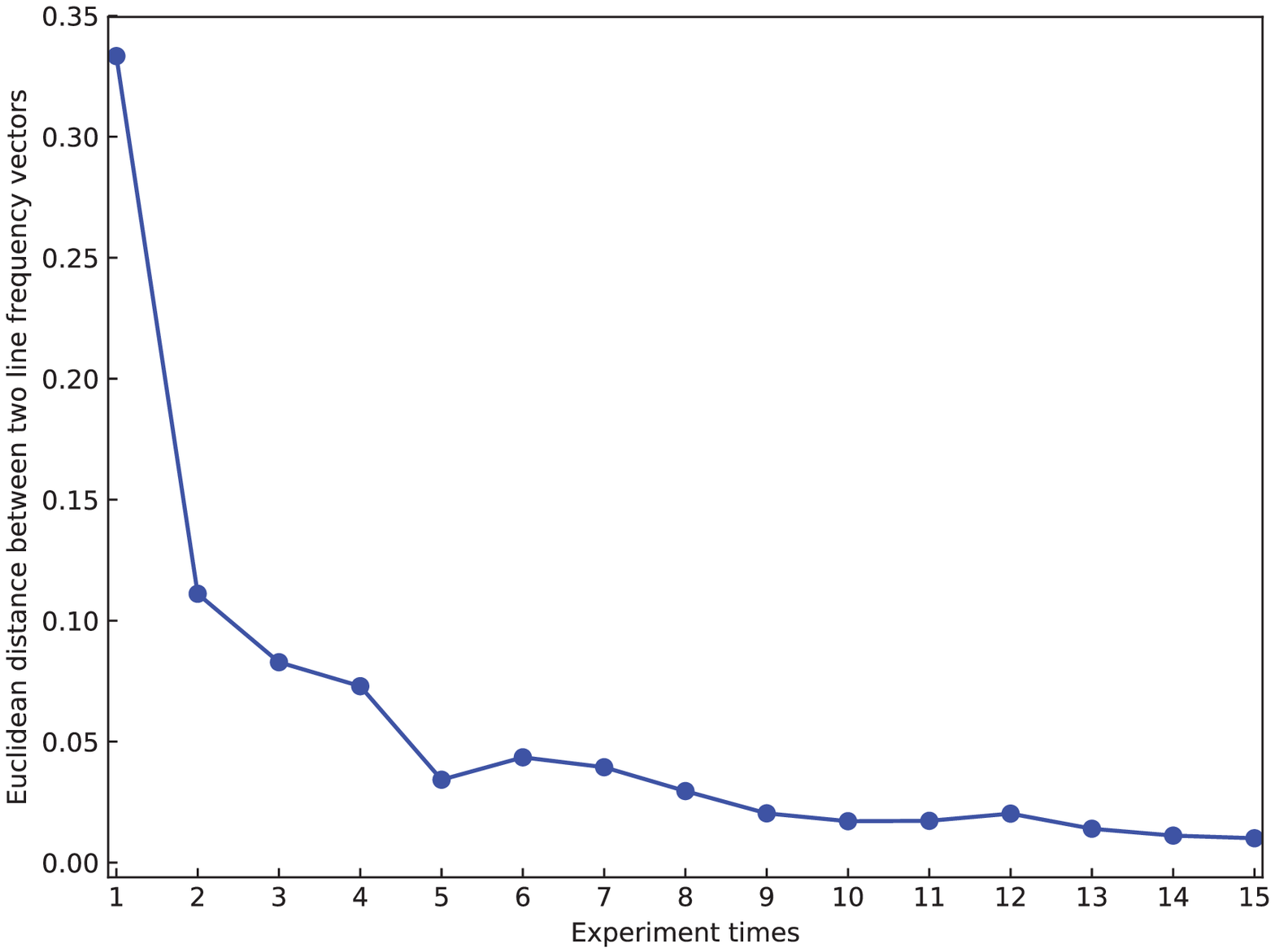}
\end{minipage}
}
\subfigure[9 lines with the highest occurrence frequencies]{
\begin{minipage}[b]{0.35\textwidth}
\includegraphics[width=1\textwidth]{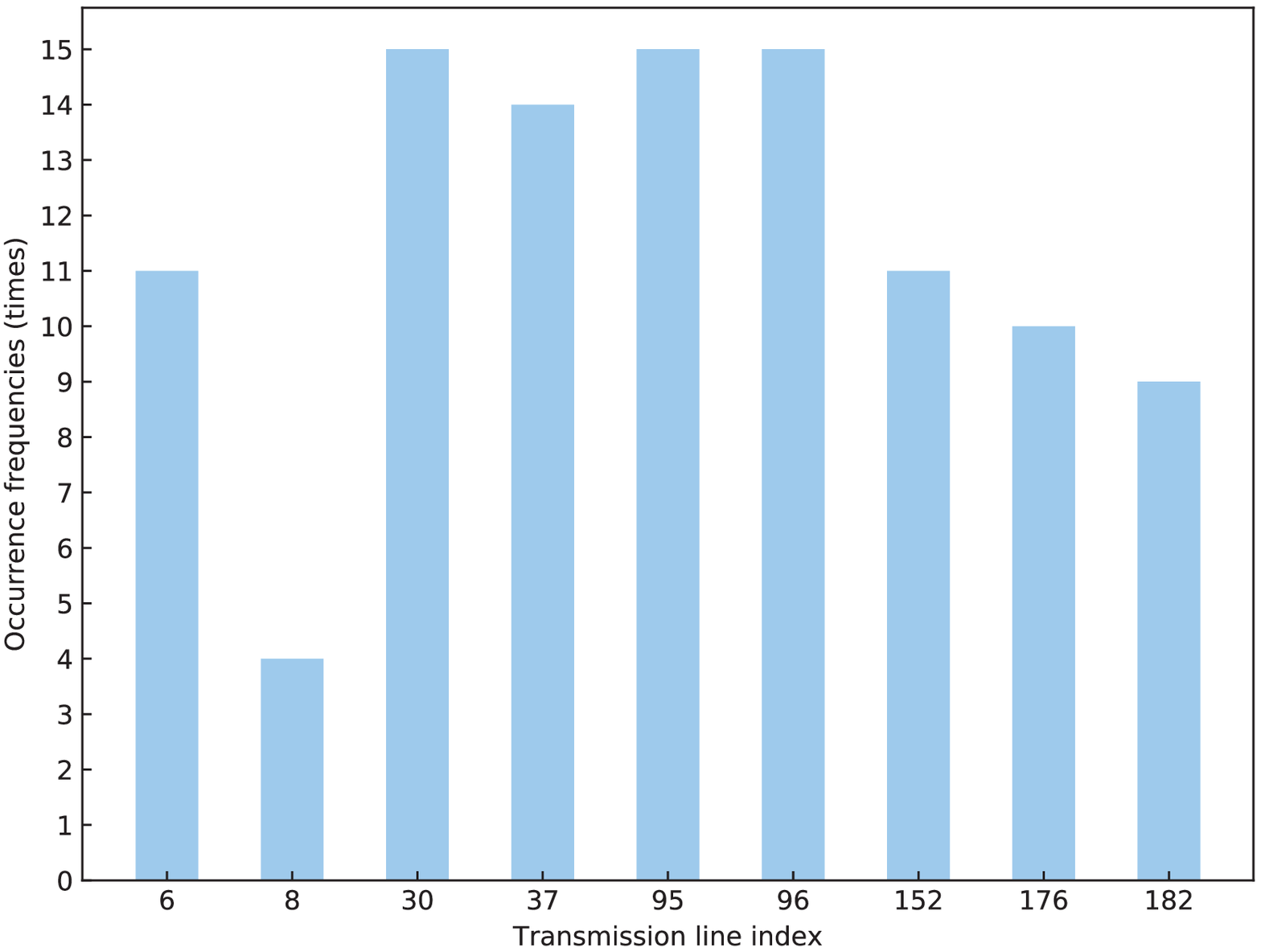}
\end{minipage}
}\\
\subfigure[Defense performance comparison]{
\begin{minipage}[b]{0.35\textwidth}
\includegraphics[width=1\textwidth]{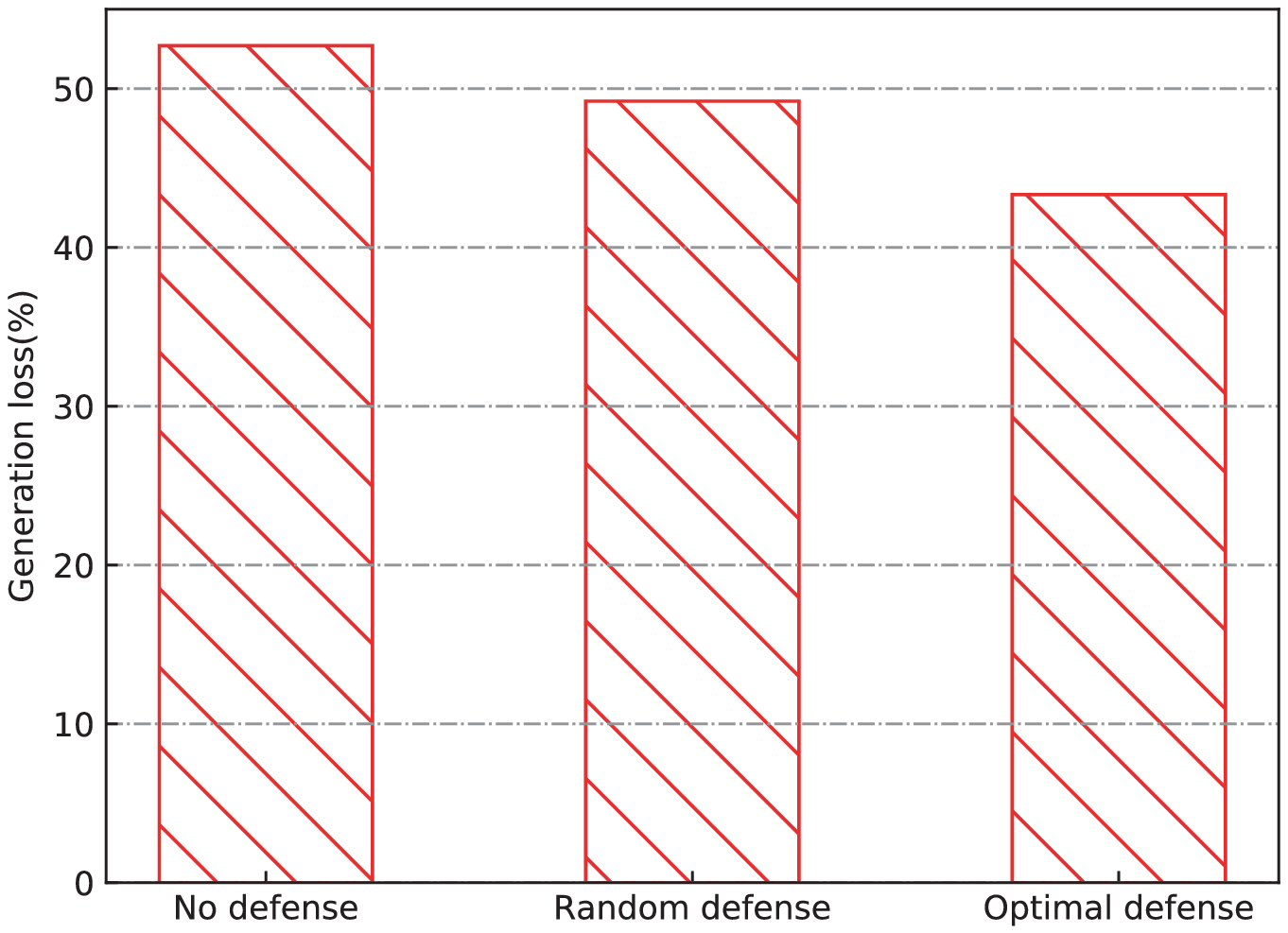}
\end{minipage}
}
\subfigure[Generation output comparison]{
\begin{minipage}[b]{0.356\textwidth}
\includegraphics[width=1\textwidth]{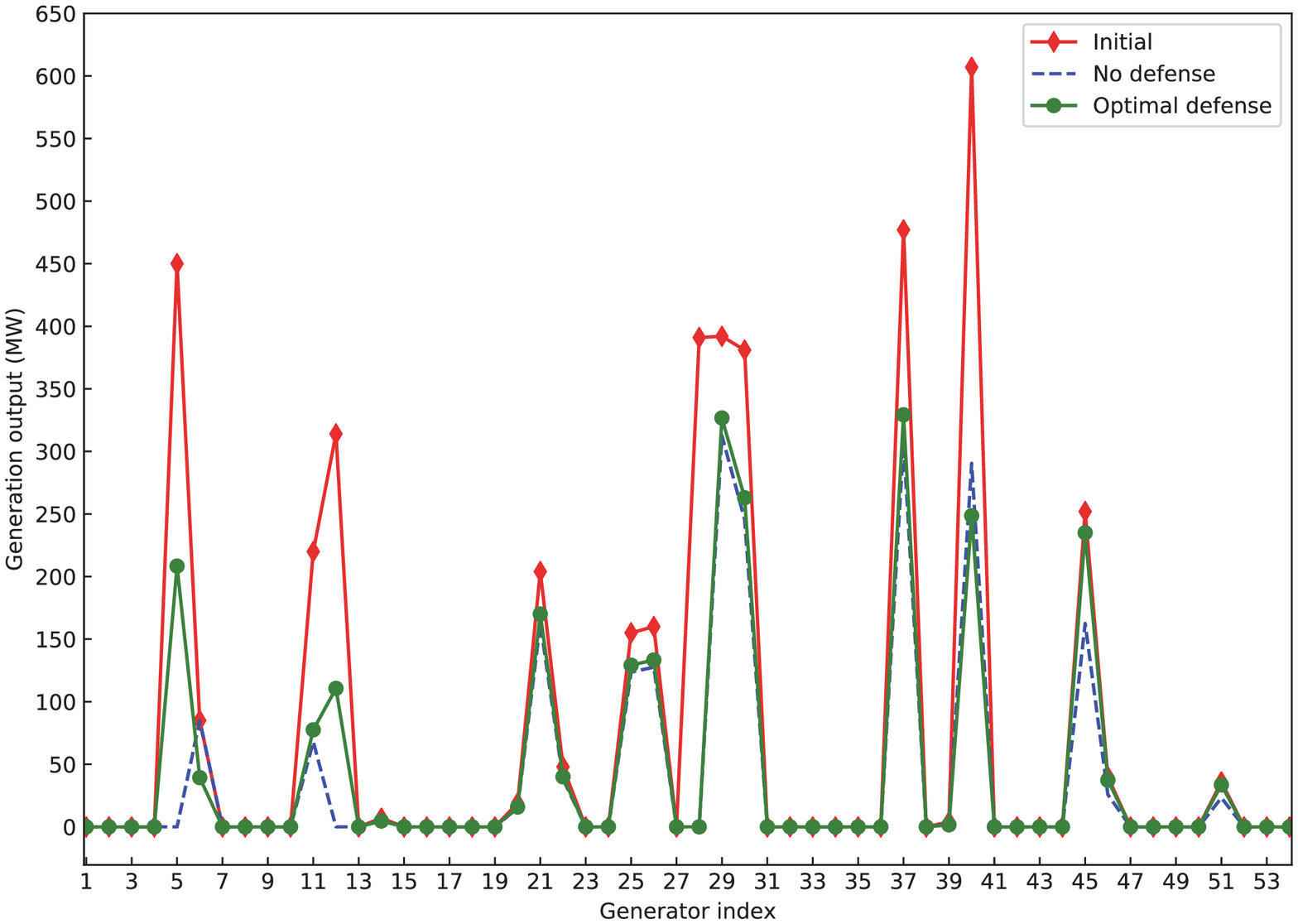}
\end{minipage}
}
\caption{The performance of the designed defense strategy.} \label{fig_7}
\end{figure*}

The results related to the designed defense strategy are shown in Fig.~\ref{fig_7}. It can be seen that the frequencies of attack lines are become more and more stable with the increase of experiment number. Here, Euclidean distance between two line frequency vectors is used as the stability metric. When 9 lines with the highest occurrence frequencies are not changed, the experiment process can be terminated and the optimal defense line set can be determined as shown in Fig.~\ref{fig_7}(b). In Figs.~\ref{fig_7}(c) and (d), the designed defense strategy is effective in reducing generation loss when compared with other schemes. To be specific, the defense strategy under coordinated multistage attacks can reduce generation loss by 17.79\% and 6.62\% compared with no defense scheme and random defense scheme (this scheme use the defense set \{0, 21, 42, 63, 84, 105, 126, 147, 168\}), respectively.



\section{Conclusions}\label{s6}
In this paper, we proposed an algorithm based on multi-agent DRL and prioritized experience replay to identify the critical lines under coordinated multistage attacks that may initiate a cascading failure. Moreover, we designed an optimal defense strategy based on the obtained optimal attacking line sequences and the available defense resources. Simulation results showed the effectiveness of the proposed algorithm and the designed strategy. This work can help the power system operator to deploy the limited defense resources optimally and mitigate the impact caused by coordinated multistage attacks.


\begin{thebibliography}{1}

\bibitem{Cheng2016}
M. Cheng, M. Crow, and Q. Ye, ``A game theory approach to vulnerability analysis: Integrating power flows with topological analysis," \emph{Electrical Power and Energy Systems}, vol. 82, pp. 29-36, 2016.


\bibitem{Bialek2016}
J. Bialek, et. al., ``Benchmarking and validation of cascading failure analysis tools," \emph{IEEE Trans. Power Systems}, vol. 31, no. 6, pp. 4887-4900, 2016.

\bibitem{Jiang2019}
H. Jiang, Z. Wang, and H. He, ``An evolutionary computation approach for smart grid cascading failure vulnerability analysis," \emph{Proc. of SSCI}, 2019.


\bibitem{Yan2015}
J. Yan, Y. Tang, Y. Zhu, H. He, Y. Sun, ``Smart grid vulnerability under cascade-based sequential line-switching attacks," \emph{Proc. of GLOBECOM}, 2015.

\bibitem{LiL2020}
L. Li, H. Wu, Y. Song, and Y. Liu, ``A state-failure-network method to identify critical components in power systems," \emph{Electric Power Systems Research}, vol. 181, pp. 106192:1-10, 2020.

\bibitem{Eppstein2012}
M. Eppstein and P. Hines, ``A ``random chemistry" algorithm for identifying collections of multiple contingencies that initiate cascading failure," \emph{IEEE Trans. Power Systems}, vol. 27, no. 3, pp. 1698-1705, 2012.


\bibitem{Chu2017}
C. Chu and H. Iu, ``Complex networks theory for modern smart grid applications: a survey," \emph{IEEE Journal on Emerging and Selected Topics in Circuits and Systems}, vol. 7, no. 2, pp. 177-191, 2017.


\bibitem{Farraj2016}
A. Farraj, E. Hammad, A. Daoud, and D. Kundur, ``A game-theoretic analysis of cyber switching attacks and mitigation in smart grid systems," \emph{IEEE Trans. Smart Grid}, vol. 7, no. 4, pp. 1846-1855, 2016.

\bibitem{Xiang2017}
Y. Xiang and L. Wang, ¡°A game-theoretic study of load redistribution attack and defense in power systems," \emph{Electric Power Systems Research}, vol. 151, pp. 12-25, 2017.

\bibitem{Arroyo2010}
J. Arroyo, ``Bilevel programming applied to power system vulnerability analysis under multiple contingencies," \emph{IET Generation, Transmission \& Distribution}, vol. 4, no. 2, pp. 178-190, 2010.

\bibitem{TianM2019}
M. Tian, M. Cui, Z. Dong, X. Wang, S. Yin, L. Zhao, ``Multilevel programming-based coordinated cyber physical attacks and countermeasures in smart grid," \emph{IEEE Access}, vol. 7, pp. 9836-9847, 2019.

\bibitem{Xiang2019}
Y. Xiang and L. Wang, ``An improved defender-attacker-defender model for transmission line defense considering offensive resource uncertainties," \emph{IEEE Trans. Smart Grid}, vol. 10, no. 3, pp. 2534-2546, 2019.


\bibitem{Ni2019}
Z. Ni and S. Paul, ``A multistage game in smart grid security: a reinforcement learning solution," \emph{IEEE Trans. Neural Networks and Learning Systems}, vol. 30, no. 9, pp. 2684-2695, 2019.

\bibitem{Xiang2020}
Y. Xiang, X. Zhang, D. Shi, R. Diao, and Z. Wang, ``Robust optimization for transmission defense against multi-period attacks with uncertainties," \emph{International Journal of Electrical Power \& Energy Systems}, vol. 121, pp. 106154:1-13, 2020.

\bibitem{Yan2017}
J. Yan, H. He, X. Zhong, and Y. Tang, ``Q-learning-based vulnerability analysis of smart grid against sequential topology attacks," \emph{IEEE Trans. Information Forensics and Security}, vol. 12, no. 1, pp. 200-210, Jan. 2017.

\bibitem{Wei2018}
L. Wei, A.I. Sarwat, W. Saad, and S. Biswas, ``Stochastic games for power grid protection against coordinated cyber-physical attacks," \emph{IEEE Trans. Smart Grid}, vol. 9, no. 2, pp. 684-694, 2018.




\bibitem{Mnih2015}
V. Mnih, et. al., ``Human-level control through deep reinforcement learning," \emph{Nature}, vol. 518, pp. 529-541, 2015.

\bibitem{Zhangzhidong2020}
Z. Zhang, D. Zhang, and R. Qiu, ``Deep reinforcement learning for power system: an overview," \emph{CSEE Journal of Power and Energy Systems}, vol. 6, no. 1, pp. 213-225, 2020.

\bibitem{Littman1994}
M. Littman, ``Markov games as a framework for multi-agent reinforcement learning," \emph{Machine Learning Proceedings}, pp. 157-163, 1994.


\bibitem{LiangTSG2020}
L. Yu, Y. Sun, Z. Xu, C. Shen, D. Yue, T. Jiang, X. Guan, Multi-agent deep reinforcement learning for HVAC control in commercial buildings, \emph{IEEE Trans. Smart Grid}, DOI: 10.1109/TSG.2020.3011739, 2020.


\bibitem{Schaul2016}
T. Schaul, J. Quan, I. Antonoglou and D. Silver, ``Prioritized experience replay," \emph{Proc. of ICLR}, 2016.


\bibitem{Paul2017}
S. Paul and Z. Ni, ``Vulnerability analysis for simultaneous attack in smart grid security," \emph{Proc. of ISGT}, 2017.



\bibitem{Iqbal2019}
S. Iqbal and F. Sha, ``Actor-attention-critic for multi-agent reinforcement learning," \emph{Proc. of ICML}, 2019.


\bibitem{Yang2019}
Y. Yang, Q. Zhai, X. Guan, and L. Wu, ``Cascading failure propogation with dynamic load variations: a simulation and mitigation framework," \emph{Proc. of GESGM}, 2019.


\bibitem{Oh2016}
J. Oh, V. Chockalingam, S. Singh, and H. Lee, ``Control of memory, active perception, and action in minecraft," https://arxiv.org/abs/1605.09128, 2016.



\end{thebibliography}
\end{document}